\documentclass[prd,superscriptaddress,footinbib,twocolumn,floatfix]{revtex4}
\usepackage{amssymb}
\usepackage{graphics}

\makeatletter
\renewcommand{\@makecaption}[2]{
  \vskip\abovecaptionskip
  \sbox\@tempboxa{\small\sf #1: #2}%
  \ifdim \wd\@tempboxa >\hsize
  \small\sf #1: #2\par
  \else
    \global \@minipagefalse
    \hb@xt@\hsize{\hfil\box\@tempboxa\hfil}%
  \fi
  \vskip\belowcaptionskip}
\makeatother

\def\ba{\begin{eqnarray}}
\def\ea{\end{eqnarray}}

\def\bar{\overline}

\def\Dslash{\,\,{\raise.15ex\hbox{/}\mkern-12mu D}}
\def\Dbarslash{\,\,{\raise.15ex\hbox{/}\mkern-12mu {\bar D}}}
\def\delslash{\,\,{\raise.15ex\hbox{/}\mkern-9mu \partial}}
\def\delbarslash{\,\,{\raise.15ex\hbox{/}\mkern-9mu {\bar\partial}}}
\def\pslash{\,\,{\raise.15ex\hbox{/}\mkern-9mu p}}
\def\calDslash{\,\,{\raise.15ex\hbox{/}\mkern-12mu {\cal D}}}

\newcommand{\Z}{{\mathbb Z}}
\newcommand{\R}{{\mathbb R}}

\def\CC{{\mathcal C}}

\def\CM{{\mathcal M}}
\def\CN{{\mathcal N}}

\def\CT{{\mathcal T}}

\newcommand{\cp}{{\mathbb{C}}{\mathbf{P}}}

\renewcommand{\bar}{\overline}

\begin{document}
\preprint{CALT 68-2885}

\title{Duality Defects}

\author{Abhijit \surname{Gadde}}
\email{abhijit@caltech.edu}
\affiliation{California Institute of Technology, Pasadena, CA 91125, USA}
\author{Sergei \surname{Gukov}}
\email{gukov@theory.caltech.edu}
\affiliation{California Institute of Technology, Pasadena, CA 91125, USA}
\affiliation{Max-Planck-Institut f\"ur Mathematik, Vivatsgasse 7, D-53111 Bonn, Germany}
\author{Pavel \surname{Putrov}}
\email{putrov@theory.caltech.edu}
\affiliation{California Institute of Technology, Pasadena, CA 91125, USA}

\begin{abstract}
We propose a unified approach to a general class of codimension-2 defects in field theories with non-trivial duality symmetries
and discuss various constructions of such ``duality defects'' in diverse dimensions.
In particular, in $d=4$ we propose a new interpretation of the Seiberg-Witten $u$-plane by ``embedding'' it in the physical space-time:
we argue that it describes a BPS configuration of two duality defects (at the monopole/dyon points)
and propose its vast generalization based on Lefschetz fibrations of 4-manifolds.
\end{abstract}


\maketitle

\newcommand{\be}{\begin{equation}}
\newcommand{\ee}{\end{equation}}

\subsection*{Introduction}

We introduce and study codimenion-2 duality defects that share certain features with surface operators, on the one hand, and with twist fields, on the other.
Surface operators are usually defined~\cite{GW} as codimension-2 singularties for various fields that often can be observed by a monodromy on a small loop around the singularity\footnote{In a four-dimensional theory, both dimension and codimension of such singularities are equal to 2; hence the name. In $d$ space-time dimensions, however, the name ``surface operator'' is a bit of a misnomer because it is the codimension (rather than dimension) that plays an important role.}.

As is well known, a good practice in quantum field theory (QFT) is to view coupling constants --- that we collectively denote by $u$ --- as background values of non-dynamical fields~\cite{Intriligator:1995au}. With this principle in mind, it is therefore natural to consider codimension-2 singularities for coupling constants as well, which can be characterized by non-trivial monodromies in the parameter space $\CM$ where $u$ take values and whose points, by definition, label inequivalent theories.
This will result in a duality defect labeled by an element $\gamma \in \Gamma = \pi_1 (\CM)$.

More generally, one can consider a field theory in $d$ dimensions with a symmetry or duality group $\Gamma$ and
introduce a codimenson-2 operator by choosing an element
\be
\gamma \in \Gamma
\ee
and requiring the theory to undergo a duality transformation by $\gamma$ around a small circle that links the support of such ``duality defect'':
$$
\resizebox{50mm}{!}{\includegraphics{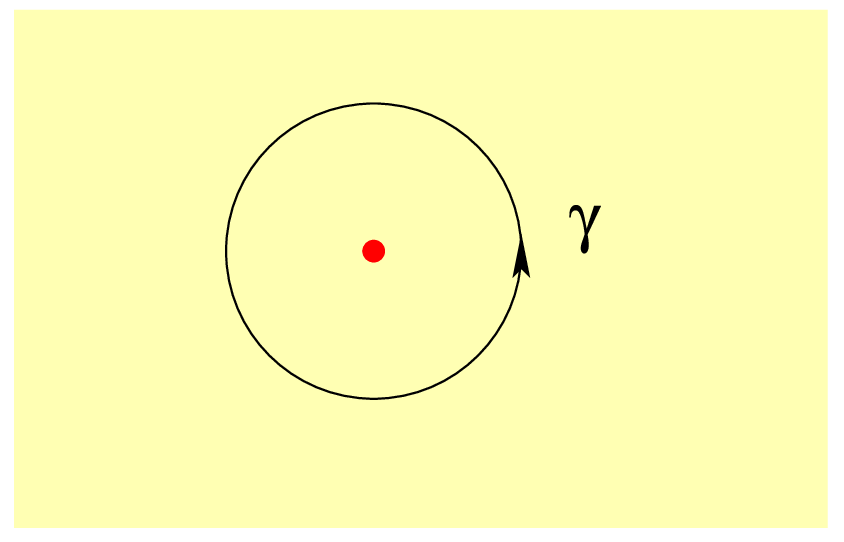}}
$$
At this level of generality, the origin of the symmetry group $\Gamma$ can be quite different, as we illustrate in numerous examples below. In particular, it can be a discrete symmetry of the 2d orbifold theory, in which case the codimension-2 duality defects are the familiar twist fields.

Much like twist fields, the codimension-2 duality defects are labeled by conjugacy classes of $\Gamma$.
Note, that even in the most familiar case of electric-magnetic duality group $\Gamma = SL(2,\Z)$
the set of conjugacy classes is rather interesting and has to do with class numbers of quadratic number fields.

While such codimension-2 duality defects at first may seem esoteric, examples are abound and, in fact, some simple cases are actually familiar.
Thus, our goal in the rest of this note will be to demystify the notion of such duality defects and to explore many examples in diverse dimensions.


\subsection*{Boundaries of duality walls}

One way to think of codimension-2 duality defects is as boundaries of duality walls
that implement duality transformations on the fields and might be more familiar to the reader.
Usually, a duality wall is defined as a codimension-1 interface which is invertible
and, therefore, gives a complete equivalence between a theory and its image under the symmetry $\gamma \in \Gamma$.
In $d=2$, where duality walls are most studied and best understood,
they are sometimes also called ``duality defects'' \cite{Frohlich:2004ef,Frohlich:2006ch,Davydov:2010rm}
which should not be confused with codimension-2 defects considered here.
In other words, being codimension-1 objects, duality walls are defect lines in $d=2$
that can end on codimension-2 objects, {\it i.e.} points in the two-dimensional space-time.
The latter are our main subject, while the former are better understood and can serve as a useful tool.
In general, a duality wall in $d$ space-time dimensions has world-volume of dimension $d-1$
and is labeled by an element of the duality group $\Gamma$.

\begin{figure}
\resizebox{60mm}{!}{\includegraphics{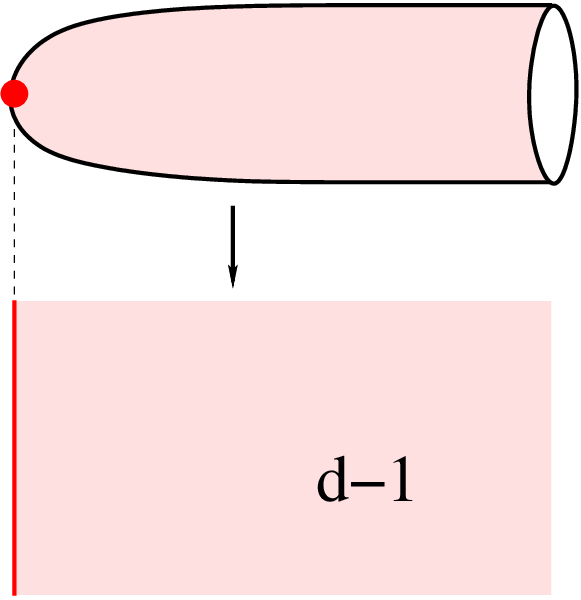}}
\caption{\label{fig:dcigar}\small Identifying $D \cong \R^2$ with a cigar offers another perspective on a codimension-2 duality defect: it corresponds to a boundary condition in $(d-1)$-dimensional theory obtained by a compactification on $S^1$ with a duality twist by $\gamma \in \Gamma$.}
\end{figure}

Let $D$ be the normal bundle ({\it i.e.} the transverse space) of the codimension-2 duality defect,
which in most of our discussion we take to be $D \cong \R^2$.
And consider a $d$-dimensional theory on $\R^2 \times M_{d-2}$.
Then, in order to define a duality defect supported on $p \times M_{d-2}$ we consider the theory
in space-time
\be
(\R^2 \times M_{d-2}) \setminus (p \times M_{d-2}) \; \cong \; S^1 \times \R \times M_{d-2}
\ee
where $p$ is the ``origin'' of $D \cong \R^2$ and we identified a punctured plane with a cylinder, $S^1 \times \R$.
Alternatively, we can take $D \cong \R^2$ to be a cigar, shown in Figure~\ref{fig:dcigar}.

Now, the nature of the codimension-2 duality defect is not that mysterious because compactifications with duality twists have been extensively studied in the
literature \cite{Kumar:1996zx,Dabholkar:1998kv,Ganor:1998ze,Kawai:2007nb,Ganor:2008hd,Ganor:2010md,Dimofte:2011jd,Ganor:2012mu,Ganor:2012ek,Gadde:2013wq}.
Namely, a $d$-dimensional theory on a circle with a non-trivial duality twist $\gamma$ gives a certain theory in $d-1$ dimensions.
Defining a boundary condition in this $(d-1)$-dimensional theory means specifying a $(d-2)$-dimensional theory $T_{d-2}$ that
describes propagating modes on the boundary and their coupling to the ``bulk theory'' in $d-1$ dimensions,
or --- if we keep the circle $S^1$ of small but finite size --- coupling to the original $d$-dimensional theory with a duality twist $\gamma$.

Hence, we can define a codimension-2 duality defect in $d$-dimensional theory by coupling to $T_{d-2}$.
This is precisely the setup of \cite{Gadde:2013wq} where boundaries of duality walls were identified
with codimension-2 defects in some simple cases, including duality defects associated with closed loops in the parameter space.
For example, a system of chiral 2d fermions coupled to 4d gauge theory generates a monodromy $\theta \to \theta + 2\pi$ and, therefore,
describes a duality defect with $\gamma = T \in SL(2,\Z)$.

Note, if the original theory in $d$ dimensions has supersymmetry and duality walls preserve half of it
(same as the theory $T_{d-1}$ defined via $S^1$ compactification with a duality twist), then typical boundary conditions
will break half of the remaining supersymmetry, which makes codimension-2 duality defects only $\frac{1}{4}$-BPS objects.

Now 
we can look for examples.
They are actually abound and exist in all dimensions and in all kinds of theories, starting with $d=2$ where codimension-2 duality defects are local, {\it i.e.} supported at points.


\subsection*{Duality defects in $d=2$}

Many interesting two-dimensional systems enjoy non-trivial duality symmetries $\Gamma \ne 1$,
which often serve as prototypes of various dualities in higher dimensions.
Moreover, in many cases, one can identify one-dimensional duality walls (= invertible line defects) that implement these dualities
and even describe codimension-2 (local) operators on which such defect lines can terminate.

The most elementary example of such system is probably 2d Ising model
that has the well-known Kramers-Wannier duality and serves as a prototype of many RCFT models.
It has three duality line operators/defects \cite{Frohlich:2004ef}: the trivial (identity) line and two non-trivial line defects,
one of which is the familiar line of antiferromagnetic couplings that can end on the insertion of disorder field.
A lot more interesting is the line defect that implements the Kramers-Wannier duality, which is fairly well understood.
It would be interesting to work out a codimension-2 defect on which this line defect can terminate;
based on the operator algebra, roughly speaking, it should be a ``square root'' of the disorder operator.
Similar duality line defects exist in other rational conformal field theories \cite{Frohlich:2006ch,Davydov:2010rm};
their end-points are precisely the codimension-2 defects of our interest here.

Another large class of 2d models with interesting duality symmetries includes variants of sigma-models,
which are attractive due to their close connection with geometry.
Thus, simple geometric symmetries of a sigma-model with target space $X$ are represented by
\be
\Delta \subset X \times X
\label{DXX}
\ee
where $\Delta$ is the graph of the isometry $\gamma$ or the so-called correspondence.
In $d=2$ space-time of the sigma-model, (\ref{DXX}) provides an identification of the fields
on the two sides of the line defect that implements such a symmetry transformation.
For example, the diagonal in $X \times X$ corresponds to the trivial (sometimes called ``invisible'') line operator.
Other concrete examples of duality line defects in simple sigma-models with various amount of
supersymmetry (including non-supersymmetric ones) can be found {\it e.g.} in \cite{Kapustin:2010zc}.

In the context of topological sigma-models, a generalization of such symmetries is usually expressed
in terms of autoequivalences of the category $\CC (X)$ of boundary conditions:
\be
\Gamma = \text{Autoeq} (\CC(X))
\label{AuteqG}
\ee
Namely, in the topological $A$-model $\CC (X)$ is the (derived) Fukaya category of $X$,
whereas in the $B$-model $\CC (X)$ is the derived category of coherent sheaves on $X$.
The duality group $\text{Autoeq} (\CC(X))$ includes shifts of the $B$-field,
geometric symmetries described above, as well as more sophisticated dualities.
When the topological theory is obtained by a (topological) twist of a supersymmetric sigma-model,
$\Gamma$ often extends to a symmetry of the physical theory.

Furthermore, such symmetries of sigma-models 
often can be realized as monodromies in a suitable parameter space $\CM$, see {\it e.g.} \cite{GW}.
Then, one can define codimension-1 duality walls by allowing the parameters $u \in \CM$
to be a function of the space-time coordinate $x$, which changes from a given value of $u$ at $x = -\infty$
to its image under a duality symmetry, $\gamma (u)$, at $x = + \infty$.
For concrete examples see {\it e.g.} \cite{GW}
where monodromies in the parameter space $\CM$ of 2d sigma-models were identified with familiar line operators.
End-points of such line operators are codimension-2 operators that we interested in.

For instance, the K\"ahler moduli space of the $\CN=(4,4)$ supersymmetric sigma-model with target space $X = T^* \cp^1$ is
a copy of complex plane with two special points, similar to Figure~\ref{fig:uplane};
the two special points are the conifold and orbifold points of the 2d sigma-model.
Identifying this moduli space with the space-time domain $D$
we obtain a BPS configuration of two codimension-2 duality defects
associated with the duality twists around the conifold and orbifold point.

It turns out that even more sophisticated dualities of sigma-models, such as T-dualities, Fourier-Mukai transforms,
and mirror symmetry correspond to codimension-1 duality walls \cite{Kawai:2007nb,Ganor:2012ek,Bachas:2012bj,Sarkissian:2008dq,Sarkissian:2010bw}.
It would be interesting to study local (codimension-2) operators on which these line defects can end.
In some cases, such sigma-models, T-duality walls, and their end-point defects
admit a ``lift'' to $d=3$ and $d=4$ theories that we consider next.
In particular, the $T^* \cp^1$ sigma-model with its codimension-2 duality defects
is a special case of $\CN=(4,4)$ sigma-models based on Hitchin moduli space that arise
via dimensional reduction from 4d $\CN=2$ theories discussed below.


\subsection*{Duality defects in $d=3$}

In $d=3$, the simplest examples of dualities and codimension-1 walls which realize them can be found in Chern-Simons gauge theory (with and without supersymmetry).
Such dualities include the famous level-rank duality, a large network of quiver Chern-Simons theories related by ``3d Kirby moves'' \cite{Gadde:2013sca},
as well as holographic ``lifts'' of some dualities in two-dimensional CFTs described earlier.

A special case of the latter is a 3d lift of the T-duality, which is non-trivial even in the basic case of the Abelian theory
with gauge group $U(1)$ at level $k=2N$.
As shown in \cite{Fuchs:2002cm,Kapustin:2010if}, this theory admits an invertible topological defect (a.k.a. ``duality surface operator'')
which implements a T-duality in the free boson CFT on the boundary.
Note, even though this transformation has no obvious interpretation as a monodromy in some parameter space, it can still be realized by a codimension-1 duality wall.
This duality wall can end on a codimension-2 defect, which in $d=3$ is a line operator, in fact, a particular Wilson line.

In 3d supersymmetric theories, there are also duality walls associated with closed loops in the parameter space,
see {\it e.g.} \cite{Gadde:2013wq} for discussion of half-BPS walls in $\CN=2$ theories.
In this case, however, the corresponding 2d theory on the wall carries $\CN=(0,2)$ supersymmetry
and, as such, does not admit BPS boundary conditions.
In view of this, it would be interesting to explore codimension-2 BPS configurations in three-dimensional theories,
especially those which can not be obtained by a lift from two dimensions.


\subsection*{Duality defects in $d=4$}

The case of $d=4$ is by far the most interesting and mysterious, mainly because dualities of 4d theories are much less trivial than in $d=2$ or $d=3$.
Nevertheless, we claim that codimension-2 duality defects not only exist in 4d theories, but can be BPS, {\it i.e.} preserve a fraction of supersymmetry.

We explain this in the context of 4d gauge theories with $\CN=2$ and $\CN=4$ supersymmetry.
(Generalization to other theories --- in particular, without SUSY --- should follow similar lines.)
Also, one can explore UV or IR regimes of the theory, and we will try to cover both considering them in parallel.

\begin{figure}
\resizebox{60mm}{!}{\includegraphics{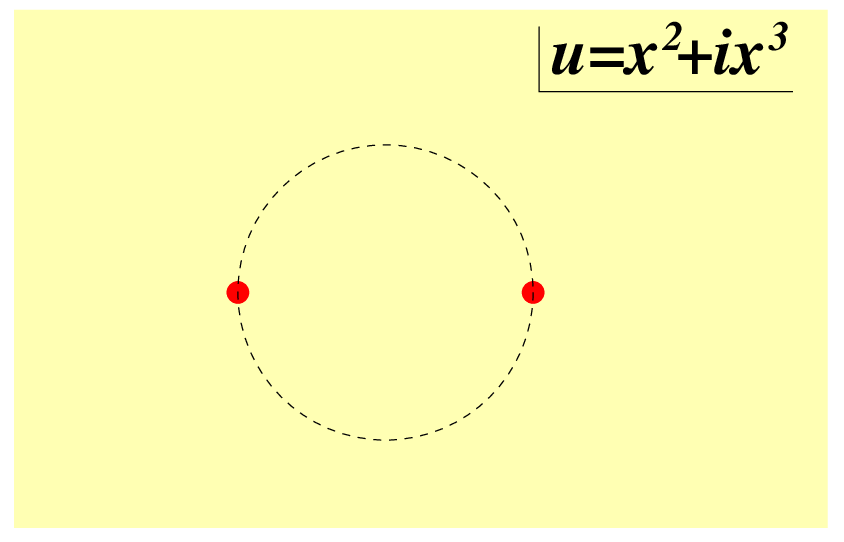}}
\caption{\label{fig:uplane}\small The simplest holomorphic map $D \to \CM_C$ is given by identifying the space-time $(x^2,x^3)$ plane with the $u$-plane.}
\end{figure}

At low energies, vacua of 4d gauge theories with extended supersymmetry are labeled by points in a suitable moduli space $\CM$,
the prominent examples of which are the Coulomb branch $\CM_C$ or the Higgs branch $\CM_H$.
The (holomorphic) coordinates $u \in \CM$ play the role of parameters that can undergo monodromies ({\it i.e.} traverse non-trivial loops in $\CM$)
along a small space-time circle around a duality defect.
In other words, $u$ can vary as a function of coordinates on the two-dimensional part of space-time, $D$,
thereby defining a map
\be
D \to \CM
\label{DtoM}
\ee
For simplicity, let us suppose that $D = \R^2$ is the $(x^2,x^3)$ plane and introduce a holomorphic coordinate $z = x^2 + i x^3$.
Then, the supersymmetry condition requires the map (\ref{DtoM}) to be holomorphic,
\be
\bar \partial_z u =0
\label{uzhol}
\ee
Depending on whether $\CM$ stands for the Coulomb or Higgs branch,
such solutions preserve either $\CN=(0,4)$ or $\CN=(0,2)$ supersymmetry in two remaining dimensions that we denote by $x^0$ and $x^1$.

For concreteness, let us focus on the Coulomb branch of $\CN=2$ super-Yang-Mills theory regarded as the parameter space
of the low-energy Seiberg-Witten theory \cite{Seiberg:1994rs}.
In the basic case of $SU(2)$ theory, the Coulomb branch is the famous $u$-plane (which explains our choice of notations).
The simplest solution to (\ref{uzhol}) is the holomorphic function
\be
u(z) = z
\label{uplanesol}
\ee
that simply identifies the $u$-plane with the space-time plane $(x^2,x^3)$.
Figuratively speaking, one might describe the identification (\ref{uplanesol}) of the $u$-plane with the space-time $(x^2,x^3)$ plane
by saying that ``$u$-plane is a snapshot of a BPS configuration''.

With this choice of a BPS solution, there are two special points on the $(x^2,x^3)$ plane,
namely the famous monopole and dyon condensation points, illustrated in Figure~\ref{fig:uplane}.
Since going around these points amounts to duality transformation by elements of the duality group $\Gamma (2) \subset SL(2,\Z)$,
\be
\gamma_{1} = \left(
\begin{array}{cc}
-1 & 2 \\
-2 & 3
\end{array}
\right)
\quad , \quad
\gamma_2 = \left(
\begin{array}{cc}
1 & 0 \\
-2 & 1
\end{array}
\right)
\label{ggSW}
\ee
these two points, $p_1$ and $p_2$, in the $(x^2,x^3)$ plane are precisely the locations of the codimension-2 duality defects labeled by (\ref{ggSW}).

More generally, the preimages $p_i \in D$ of the Coulomb branch discriminant locus
under the map (\ref{DtoM}) are precisely the locations of the codimension-2 duality defects
labeled by elements of the duality group $\gamma_i \in \Gamma$.
Since the discriminant locus has codimension-2 in the Coulomb branch,
a generic map (\ref{DtoM}) meets it in isolated points (whose preimages are the points $p_i$).
For example, the same identification (\ref{uplanesol}) in $\CN=2$ SQCD with $N_f$ massless quarks \cite{Seiberg:1994aj}
gives a configuration of three duality defects for $N_f=1$ and two duality defects for $N_f=2,3$.
Turning on quark masses leads to fragmentation of the duality defects and changes their positions $p_i \in D$. 

One way to show that (\ref{uplanesol}) is a BPS solution is
to consider a dimensional reduction/compactification along the directions $x^0$ and $x^1$.
Indeed, since we are interested in configurations that do not depend on these directions at all, we can take the four-dimensional space-time to be $D \times T^2$ and study the effective theory on $D$ (in the limit of small $T^2$). The resulting theory is a 2d $\CN=(4,4)$ sigma-model with the hyper-K\"ahler target space which, depending on the context, is either $\CM$ itself or a suitable enhancement thereof.
In the case of the Coulomb branch considered here, $\CM_C$ is ``enhanced'' to the Hitchin moduli space
by the vevs of Wilson lines on $T^2$. (In fact, this makes $\CM_C$ to be the base of the Hitchin fibration.)
In any case, the standard arguments in 2d sigma-model show that (\ref{uplanesol}) is indeed a supersymmetric solution.
Note, that upon this reduction to a 2d sigma-model on $D$, the duality group of the original four-dimensional theory
becomes the group of symmetries (\ref{AuteqG}) of the two-dimensional sigma-model.

\begin{figure}
\resizebox{60mm}{!}{\includegraphics{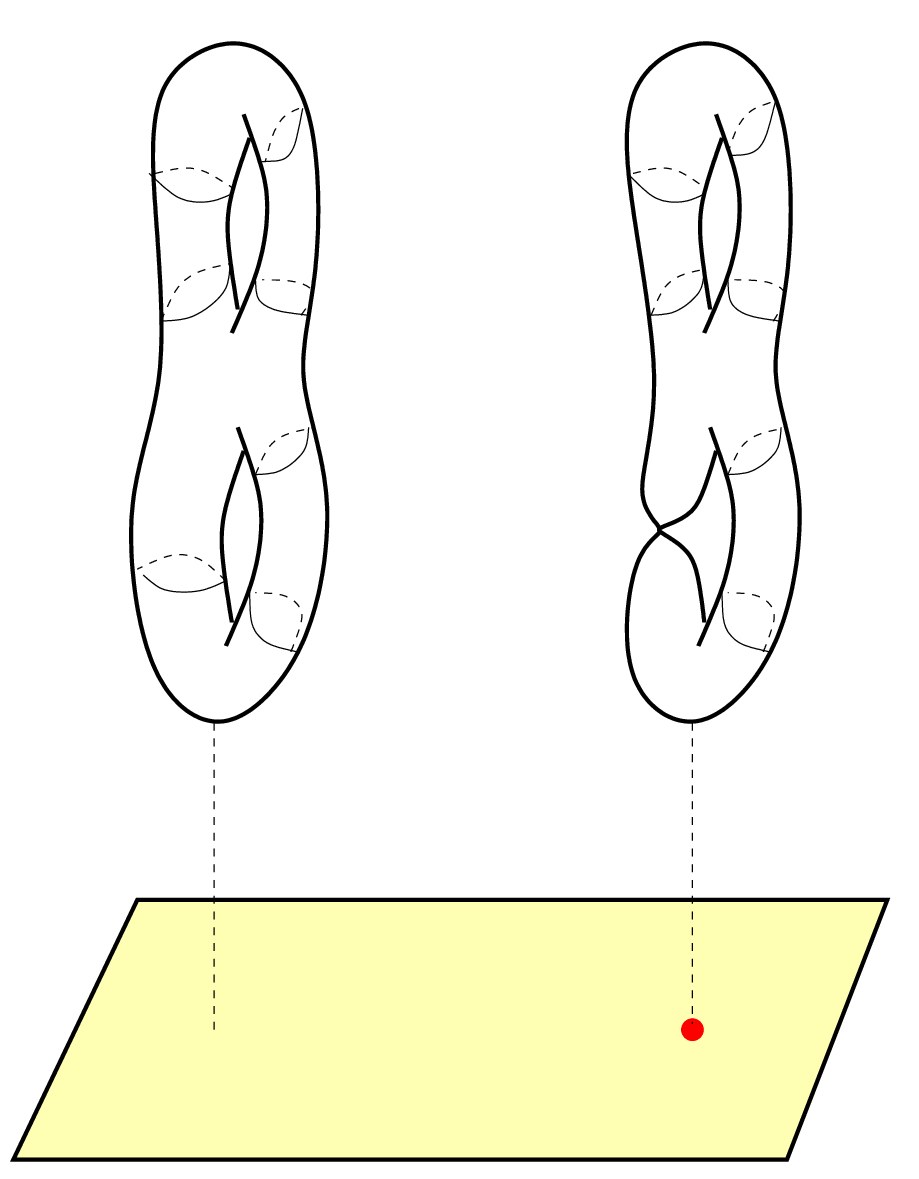}}
\caption{\label{fig:Lefschetz}\small Lefschetz fibration defines a $\frac{1}{4}$-BPS configuration of duality defects at the locations of singular fibers.}
\end{figure}

Another way to show that (\ref{uplanesol}) is a supersymmetric configuration and to generalize it to other BPS solutions
is to consider a (large) class of 4d $\CN=2$ theories produced by compactifying M-theory fivebranes on a Riemann surface $\Sigma$.
Since we treat both UV and IR descriptions in parallel, it will be convenient to use $\Sigma$ for both the IR curve (= Seiberg-Witten curve) wrapped by a single fivebrane as in \cite{Witten:1997sc}, as well as the UV curve wrapped by multiple fivebranes, as in \cite{Gaiotto:2008cd,Gaiotto:2009we}. In either case, fivebrane(s) wrapped on $\Sigma$ gives rise to a 4d $\CN=2$ gauge theory, whose couplings are complex structure parameters of the Riemann surface $\Sigma$ with values in $\CM = \CT / \text{MCG} (\Sigma)$, where
\be
\text{MCG} (\Sigma) := \text{Diff} (\Sigma) / \text{homotopy} = \pi_0 (\text{Diff} (\Sigma))
\ee
is the mapping class group of $\Sigma$ and
$\text{Diff} (\Sigma)$ is the group of orientation preserving diffeomorphisms.
For example, we have MCG$(S^2) = 0$, MCG$(T^2) = SL(2,\Z)$, {\it etc.}
Moreover, the duality group $\Gamma$ is the fundamental group of this space,
\be
\Gamma = \pi_1 (\CM) = \text{MCG} (\Sigma)
\ee
or, depending on the context, its subgroup. (The distinction will not affect the construction of duality defects below, and so will be ignored.)

We are interested in solutions, where these parameters vary over the $(x^2,x^3)$ plane or, more generally, over the base $D$
in such a way that at certain special points the fibration
\be
M_4 ~: \qquad
\begin{array}{c}
\Sigma \\
\downarrow \\
D
\end{array}
\label{Leffibr}
\ee
is singular, as illustrated in Figure~\ref{fig:Lefschetz}.
More precisely, going around these special points must produce a transformation by an element of the duality group
\be
\gamma \in \Gamma = \text{MCG} (\Sigma)
\ee
The fibration (\ref{Leffibr}) described here is precisely what in mathematical literature is called the {\it Lefschetz fibration}
and the twist around singular fiber by an element of the mapping class group is what's known as {\it Dehn twist}.
Indeed, a theorem of Lickorish states that any element $\gamma \in \text{MCG} (\Sigma)$ has a decomposition into finitely
many Dehn twists along oriented simple closed curves $\ell \subset \Sigma$.
Therefore, the slogan is:
\begin{center}
{\it Duality twist} = {\it Dehn twist}
\end{center}
Since closed curves on $\Sigma$ correspond to BPS particles (with charges given by homology class $[\ell] \in H_1 (\Sigma,\Z)$),
Dehn twists correspond to duality defects on which certain BPS particles become massless.

Moreover, half-BPS configurations defined by holomorphic maps $D\rightarrow \CM_C$
realize holomorphic Lefschetz fibrations which preserve $\CN=(0,4)$ supersymmetry in $(x^0,x^1)$ dimensions.
Indeed, since in general the curve $\Sigma$ is given by a polynomial equation $P(x,y;u)=0$
inside $\mathbb{C} \times \mathbb{C}^*$, the total space of the Lefschetz fibration (\ref{Leffibr})
is a complex surface inside a ``trivial'' Calabi-Yau 3-fold
$D\times\mathbb{C}\times\mathbb{C}^*$ (with $D\cong\mathbb{C}$) defined by the equation:
\begin{equation}
M_4 ~: \quad P (x,y;u(z))=0.
\end{equation}
Fivebranes wrapping K\"ahler 4-cycles inside Calabi-Yau 3-folds are known to preserve $\CN=(0,4)$ supersymmetry in two dimensions
(see {\it e.g.} \cite{Maldacena:1997de} and references therein).
This is a special case of a more general construction with fivebranes wrapping generic 4-manifolds
embedded as coassociative cycles inside $G_2$-holonomy spaces that leads to a large class of
two-dimensional $\CN=(0,2)$ theories $T[M_4]$ labeled by 4-manifolds~\cite{Gadde:2013sca}.

To summarize, we learn that in a large class of 4d $\CN=2$ theories
we can define many BPS duality defects
by compactifying the corresponding fivebrane system on the Lefschetz fibration (\ref{Leffibr}).
Then, each singular fiber gives a codimension-2 defect in the base of the fibration.
The resulting system of fivebrane(s) wrapped on the 4-manifold (\ref{Leffibr})
leads to the effective 2d theory $T[M_4]$ with $\CN=(0,2)$ supersymmetry,
which is enhanced to $\CN=(0,4)$ when the Lefschetz fibration is holomorphic. 

Note, constructions of $\CN=2$ theories \cite{Witten:1997sc,Gaiotto:2008cd,Gaiotto:2009we},
including the $SU(2)$ Seiberg-Witten theories discussed above,
typically involve non-compact fivebranes on $\Sigma$, whose asymptotic regions determine bare masses, couplings, {\it etc.}
Therefore, while allowing $\Sigma_u$ to vary over $D$ in (\ref{Leffibr}), it is natural to keep this boundary data fixed.
A Lefschetz fibration with these properties is called a {\it Stein manifold}.
Its boundary is a 3-manifold with an {\it open book decomposition} determined by the diffeomorphism
$\gamma : \Sigma \to \Sigma$, such that $\gamma$ 
is the identity in the neighborhood of the boundary of $\Sigma$.

Relegating a more complete account of the dictionary between geometry of Lefschetz fibrations and
physics of the corresponding BPS configurations with duality defects to a separate publication~\cite{toappear},
we conclude this section by noting that such codimension-2 configurations are ``non-local'' in the sense
that individual duality defects ``feel'' each other's presence (or, ``talk'' to each other) even when they are widely separated.
Thus, if we denote positions of duality defects (= positions of singular fibers) by $p_i \in D$,
then moving the points $p_i$ around each other on $D$ depends on the path, in particular on its homotopy type.
Perhaps this is not too surprising since, by definition, codimension-2 defects that implement non-trivial dualities are inherently strongly coupled.

This non-local structure of codimension-2 duality defects in $d=4$ is reminiscent of the modular tensor category
of codimension-2 line operators in $d=3$ TQFT or chiral algebra of codimension-2 primary vertex operators in $d=2$ CFT,
which too are labeled by configurations of points $\{ p_i \} \in D$ as in Figure~\ref{fig:uplane}.
This structure clearly has many applications, which go beyond the scope of this letter.

Finally, let us note that similar duality defects were considered in, for example, \cite{Grassi:2013kha,Martucci:2014ema}.


\subsection*{Generalizations}

Our general setup here and, in particular, the ``embedding'' of the Seiberg-Witten $u$-plane into physical space-time
admits a natural generalization.

Since singularities of the Coulomb branch are codimension-2 even for higher rank theories,
a generic holomorphic map (\ref{DtoM}) will intersect the singularity loci giving rise to defects
labeled by the elements of the associated duality group.
The special (non-generic) maps which pass through codimension-4 singularities such as Argyres-Douglas (AD) points
can be understood as collision of duality defects corresponding to mutually non-local charges.

We can also consider maps of the form (\ref{DtoM}) where the domain $D$ is not necessarily two-dimensional, but, say, four-dimensional.
Then, the generic singularities of the Coulomb branch engineer surface defects
while codimension-4 singularities such as AD points correspond to their intersections.
More generally, in a $d$-dimensional field theory, a configuration (\ref{DtoM}) with $\text{dim} (D) = k \le d$ describes
an object of codimension-$k$ (or, equivalently, of dimension $d-k$) that in supersymmetric theories can be BPS if suitable conditions are met.

\section*{Acknowledgements}

We are grateful to C.~Bachas, O.~Ganor and I.~Runkel for useful discussions.
The work of A.G. is supported in part by the John A. McCone fellowship and by DOE Grant DE-FG02-92-ER40701.
The work of S.G. is supported in part by DOE Grant DE-FG02-92ER40701.
The work of P.P. is supported in part by the Sherman Fairchild scholarship and by NSF Grant PHY-1050729.



\begin{thebibliography}{99}
\small
\parskip=0pt plus 2pt

\bibitem{GW} S.~Gukov, E.~Witten,
Current Developments in Mathematics {\bf 2006} (2008) 35-180, hep-th/0612073.

\bibitem{Intriligator:1995au}
  K.~A.~Intriligator and N.~Seiberg,
  Nucl.\ Phys.\ Proc.\ Suppl.\  {\bf 45BC}, 1 (1996)
  [hep-th/9509066].

\bibitem{Frohlich:2004ef}
  J.~Frohlich, J.~Fuchs, I.~Runkel and C.~Schweigert,
  Phys.\ Rev.\ Lett.\  {\bf 93}, 070601 (2004)
  [cond-mat/0404051].

\bibitem{Frohlich:2006ch}
  J.~Frohlich, J.~Fuchs, I.~Runkel and C.~Schweigert,
  Nucl.\ Phys.\ B {\bf 763}, 354 (2007)
  [hep-th/0607247].

\bibitem{Davydov:2010rm}
  A.~Davydov, L.~Kong and I.~Runkel,
  Adv.\ Theor.\ Math.\ Phys.\  {\bf 15} (2011)
  [arXiv:1004.4725 [hep-th]].

\bibitem{Kumar:1996zx}
  A.~Kumar and C.~Vafa,
  Phys.\ Lett.\ B {\bf 396}, 85 (1997)
  [hep-th/9611007].

\bibitem{Dabholkar:1998kv}
  A.~Dabholkar and J.~A.~Harvey,
  JHEP {\bf 9902}, 006 (1999)
  [hep-th/9809122].

\bibitem{Ganor:1998ze}
  O.~J.~Ganor,
  Nucl.\ Phys.\ B {\bf 549}, 145 (1999)
  [hep-th/9812024].

\bibitem{Kawai:2007nb}
  S.~Kawai and Y.~Sugawara,
  JHEP {\bf 0802}, 065 (2008)
  [arXiv:0711.1045 [hep-th]].

\bibitem{Ganor:2008hd}
  O.~J.~Ganor and Y.~P.~Hong,
  arXiv:0812.1213 [hep-th].

\bibitem{Ganor:2010md}
  O.~J.~Ganor, Y.~P.~Hong and H.~S.~Tan,
  JHEP {\bf 1103}, 099 (2011)
  [arXiv:1007.3749 [hep-th]].

\bibitem{Dimofte:2011jd}
  T.~Dimofte and S.~Gukov,
  JHEP {\bf 1305}, 109 (2013)
  [arXiv:1106.4550 [hep-th]].

\bibitem{Ganor:2012mu}
  O.~J.~Ganor, Y.~P.~Hong, R.~Markov and H.~S.~Tan,
  JHEP {\bf 1204}, 041 (2012)
  [arXiv:1201.2679 [hep-th]].

\bibitem{Ganor:2012ek}
  O.~J.~Ganor, S.~Jue and S.~McCurdy,
  JHEP {\bf 1302}, 017 (2013)
  [arXiv:1211.4179 [hep-th]].

\bibitem{Gadde:2013wq}
  A.~Gadde, S.~Gukov and P.~Putrov,
  arXiv:1302.0015 [hep-th].

\bibitem{Kapustin:2010zc}
  A.~Kapustin and K.~Setter,
  arXiv:1009.5999 [hep-th].

\bibitem{Bachas:2012bj}
  C.~Bachas, I.~Brunner and D.~Roggenkamp,
  JHEP {\bf 1210}, 039 (2012)
  [arXiv:1205.4647 [hep-th]].

\bibitem{Gadde:2013sca}
  A.~Gadde, S.~Gukov and P.~Putrov,
  arXiv:1306.4320 [hep-th].


\bibitem{Fuchs:2002cm}
  J.~Fuchs, I.~Runkel and C.~Schweigert,
  Nucl.\ Phys.\ B {\bf 646}, 353 (2002)
  [hep-th/0204148].

\bibitem{Kapustin:2010if}
  A.~Kapustin and N.~Saulina,
  In *Sati, Hisham (ed.) et al.: Mathematical Foundations of Quantum Field theory and Perturbative String Theory* 175-198
  [arXiv:1012.0911 [hep-th]].

\bibitem{Seiberg:1994rs}
  N.~Seiberg and E.~Witten,
  Nucl.\ Phys.\ B {\bf 426}, 19 (1994)
  [Erratum-ibid.\ B {\bf 430}, 485 (1994)]
  [hep-th/9407087].

\bibitem{Seiberg:1994aj}
  N.~Seiberg and E.~Witten,
  Nucl.\ Phys.\ B {\bf 431}, 484 (1994)
  [hep-th/9408099].

\bibitem{Witten:1997sc}
  E.~Witten,
  Nucl.\ Phys.\ B {\bf 500}, 3 (1997)
  [hep-th/9703166].

\bibitem{Gaiotto:2008cd}
  D.~Gaiotto, G.~W.~Moore and A.~Neitzke,
  Commun.\ Math.\ Phys.\  {\bf 299}, 163 (2010)
  [arXiv:0807.4723 [hep-th]].

\bibitem{Gaiotto:2009we}
  D.~Gaiotto,
  JHEP {\bf 1208}, 034 (2012)
  [arXiv:0904.2715 [hep-th]].

\bibitem{Maldacena:1997de}
  J.~M.~Maldacena, A.~Strominger and E.~Witten,
  JHEP {\bf 9712}, 002 (1997)
  [hep-th/9711053].
   
\bibitem{Sarkissian:2008dq} 
  G.~Sarkissian and C.~Schweigert,
  Nucl.\ Phys.\ B {\bf 819}, 478 (2009)
  [arXiv:0810.3159 [hep-th]].
  
\bibitem{Sarkissian:2010bw} 
  G.~Sarkissian,
  Nucl.\ Phys.\ B {\bf 846}, 338 (2011)
  [arXiv:1006.5317 [hep-th]].

\bibitem{Grassi:2013kha} 
  A.~Grassi, J.~Halverson and J.~L.~Shaneson,
  JHEP {\bf 1310}, 205 (2013)
  [arXiv:1306.1832 [hep-th]].
  
%
\bibitem{Martucci:2014ema} 
  L.~Martucci,
  JHEP {\bf 1406}, 180 (2014)
  [arXiv:1403.2530 [hep-th]].

\bibitem{toappear}
  A.~Gadde, S.~Gukov and P.~Putrov,
  to appear.


\end{thebibliography}
\end{document}